\documentclass[%
reprint,
superscriptaddress,
amsmath,amssymb,
aps,
]{revtex4-2}

\usepackage{graphicx}
\usepackage{dcolumn}
\usepackage{bm}

\usepackage{color}

\begin{document}


\title{First Detection of sub-PeV Diffuse Gamma Rays from the Galactic Disk: \\ Evidence for Ubiquitous Galactic Cosmic Rays beyond PeV Energies}

\author{M.~Amenomori}
\affiliation{Department of Physics, Hirosaki University, Hirosaki 036-8561, Japan }
\author{Y.~W.~Bao}
\affiliation{School of Astronomy and Space Science, Nanjing University, Nanjing 210093, China }
\author{X.~J.~Bi}
\affiliation{Key Laboratory of Particle Astrophysics, Institute of High Energy Physics, Chinese Academy of Sciences, Beijing 100049, China }
\author{D.~Chen$^{\S}$}
\affiliation{National Astronomical Observatories, Chinese Academy of Sciences, Beijing 100012, China }
\author{T.~L.~Chen}
\affiliation{Physics Department of Science School, Tibet University, Lhasa 850000, China }
\author{W.~Y.~Chen}
\affiliation{Key Laboratory of Particle Astrophysics, Institute of High Energy Physics, Chinese Academy of Sciences, Beijing 100049, China }
\author{Xu~Chen}
\affiliation{Key Laboratory of Particle Astrophysics, Institute of High Energy Physics, Chinese Academy of Sciences, Beijing 100049, China }
\author{Y.~Chen}
\affiliation{School of Astronomy and Space Science, Nanjing University, Nanjing 210093, China }
\author{Cirennima}
\affiliation{Physics Department of Science School, Tibet University, Lhasa 850000, China }
\author{S.~W.~Cui}
\affiliation{Department of Physics, Hebei Normal University, Shijiazhuang 050016, China }
\author{Danzengluobu}
\affiliation{Physics Department of Science School, Tibet University, Lhasa 850000, China }
\author{L.~K.~Ding}
\affiliation{Key Laboratory of Particle Astrophysics, Institute of High Energy Physics, Chinese Academy of Sciences, Beijing 100049, China }
\author{J.~H.~Fang}
\affiliation{Key Laboratory of Particle Astrophysics, Institute of High Energy Physics, Chinese Academy of Sciences, Beijing 100049, China }
\affiliation{University of Chinese Academy of Sciences, Beijing 100049, China }
\author{K.~Fang}
\affiliation{Key Laboratory of Particle Astrophysics, Institute of High Energy Physics, Chinese Academy of Sciences, Beijing 100049, China }
\author{C.~F.~Feng}
\affiliation{Institute of Frontier and Interdisciplinary Science and Key Laboratory of Particle Physics and Particle Irradiation (MOE), Shandong University, Qingdao 266237, China }
\author{Zhaoyang~Feng}
\affiliation{Key Laboratory of Particle Astrophysics, Institute of High Energy Physics, Chinese Academy of Sciences, Beijing 100049, China }
\author{Z.~Y.~Feng}
\affiliation{Institute of Modern Physics, SouthWest Jiaotong University, Chengdu 610031, China }
\author{Qi~Gao}
\affiliation{Physics Department of Science School, Tibet University, Lhasa 850000, China }
\author{Q.~B.~Gou}
\affiliation{Key Laboratory of Particle Astrophysics, Institute of High Energy Physics, Chinese Academy of Sciences, Beijing 100049, China }
\author{Y.~Q.~Guo}
\affiliation{Key Laboratory of Particle Astrophysics, Institute of High Energy Physics, Chinese Academy of Sciences, Beijing 100049, China }
\author{Y.~Y.~Guo}
\affiliation{Key Laboratory of Particle Astrophysics, Institute of High Energy Physics, Chinese Academy of Sciences, Beijing 100049, China }
\author{H.~H.~He}
\affiliation{Key Laboratory of Particle Astrophysics, Institute of High Energy Physics, Chinese Academy of Sciences, Beijing 100049, China }
\author{Z.~T.~He}
\affiliation{Department of Physics, Hebei Normal University, Shijiazhuang 050016, China }
\author{K.~Hibino}
\affiliation{Faculty of Engineering, Kanagawa University, Yokohama 221-8686, Japan }
\author{N.~Hotta}
\affiliation{Faculty of Education, Utsunomiya University, Utsunomiya 321-8505, Japan }
\author{Haibing~Hu}
\affiliation{Physics Department of Science School, Tibet University, Lhasa 850000, China }
\author{H.~B.~Hu}
\affiliation{Key Laboratory of Particle Astrophysics, Institute of High Energy Physics, Chinese Academy of Sciences, Beijing 100049, China }
\author{J.~Huang$^{\ddagger}$}
\affiliation{Key Laboratory of Particle Astrophysics, Institute of High Energy Physics, Chinese Academy of Sciences, Beijing 100049, China }
\author{H.~Y.~Jia}
\affiliation{Institute of Modern Physics, SouthWest Jiaotong University, Chengdu 610031, China }
\author{L.~Jiang}
\affiliation{Key Laboratory of Particle Astrophysics, Institute of High Energy Physics, Chinese Academy of Sciences, Beijing 100049, China }
\author{H.~B.~Jin}
\affiliation{National Astronomical Observatories, Chinese Academy of Sciences, Beijing 100012, China }
\author{K.~Kasahara}
\affiliation{Faculty of Systems Engineering, Shibaura Institute of Technology, Omiya 330-8570, Japan }
\author{Y.~Katayose}
\affiliation{Faculty of Engineering, Yokohama National University, Yokohama 240-8501, Japan }
\author{C.~Kato}
\affiliation{Department of Physics, Shinshu University, Matsumoto 390-8621, Japan }
\author{S.~Kato}
\affiliation{Institute for Cosmic Ray Research, University of Tokyo, Kashiwa 277-8582, Japan }
\author{K.~Kawata$^{*}$}
\affiliation{Institute for Cosmic Ray Research, University of Tokyo, Kashiwa 277-8582, Japan }
\author{W.~Kihara}
\affiliation{Department of Physics, Shinshu University, Matsumoto 390-8621, Japan }
\author{Y.~Ko}
\affiliation{Department of Physics, Shinshu University, Matsumoto 390-8621, Japan }
\author{M.~Kozai}
\affiliation{Institute of Space and Astronautical Science, Japan Aerospace Exploration Agency (ISAS/JAXA), Sagamihara 252-5210, Japan}
\author{Labaciren}
\affiliation{Physics Department of Science School, Tibet University, Lhasa 850000, China }
\author{G.~M.~Le}
\affiliation{National Center for Space Weather, China Meteorological Administration, Beijing 100081, China }
\author{A.~F.~Li}
\affiliation{School of Information Science and Engineering, Shandong Agriculture University, Taian 271018, China }
\affiliation{Institute of Frontier and Interdisciplinary Science and Key Laboratory of Particle Physics and Particle Irradiation (MOE), Shandong University, Qingdao 266237, China }
\affiliation{Key Laboratory of Particle Astrophysics, Institute of High Energy Physics, Chinese Academy of Sciences, Beijing 100049, China }
\author{H.~J.~Li}
\affiliation{Physics Department of Science School, Tibet University, Lhasa 850000, China }
\author{W.~J.~Li}
\affiliation{Key Laboratory of Particle Astrophysics, Institute of High Energy Physics, Chinese Academy of Sciences, Beijing 100049, China }
\affiliation{Institute of Modern Physics, SouthWest Jiaotong University, Chengdu 610031, China }
\author{Y.~H.~Lin}
\affiliation{Key Laboratory of Particle Astrophysics, Institute of High Energy Physics, Chinese Academy of Sciences, Beijing 100049, China }
\affiliation{University of Chinese Academy of Sciences, Beijing 100049, China }
\author{B.~Liu}
\affiliation{Department of Astronomy, School of Physical Sciences, University of Science and Technology of China, Hefei, Anhui 230026, China }
\author{C.~Liu}
\affiliation{Key Laboratory of Particle Astrophysics, Institute of High Energy Physics, Chinese Academy of Sciences, Beijing 100049, China }
\author{J.~S.~Liu}
\affiliation{Key Laboratory of Particle Astrophysics, Institute of High Energy Physics, Chinese Academy of Sciences, Beijing 100049, China }
\author{M.~Y.~Liu}
\affiliation{Physics Department of Science School, Tibet University, Lhasa 850000, China }
\author{W.~Liu}
\affiliation{Key Laboratory of Particle Astrophysics, Institute of High Energy Physics, Chinese Academy of Sciences, Beijing 100049, China }
\author{Y.-Q.~Lou}
\affiliation{Department of Physics and Tsinghua Centre for Astrophysics (THCA), Tsinghua University, Beijing 100084, China }
\affiliation{Tsinghua University-National Astronomical Observatories of China (NAOC) Joint Research Center for Astrophysics, Tsinghua University, Beijing 100084, China }
\affiliation{Department of Astronomy, Tsinghua University, Beijing 100084, China }
\author{H.~Lu}
\affiliation{Key Laboratory of Particle Astrophysics, Institute of High Energy Physics, Chinese Academy of Sciences, Beijing 100049, China }
\author{X.~R.~Meng}
\affiliation{Physics Department of Science School, Tibet University, Lhasa 850000, China }
\author{K.~Munakata}
\affiliation{Department of Physics, Shinshu University, Matsumoto 390-8621, Japan }
\author{H.~Nakada}
\affiliation{Faculty of Engineering, Yokohama National University, Yokohama 240-8501, Japan }
\author{Y.~Nakamura}
\affiliation{Key Laboratory of Particle Astrophysics, Institute of High Energy Physics, Chinese Academy of Sciences, Beijing 100049, China }
\author{H.~Nanjo}
\affiliation{Department of Physics, Hirosaki University, Hirosaki 036-8561, Japan }
\author{M.~Nishizawa}
\affiliation{National Institute of Informatics, Tokyo 101-8430, Japan }
\author{M.~Ohnishi}
\affiliation{Institute for Cosmic Ray Research, University of Tokyo, Kashiwa 277-8582, Japan }
\author{T.~Ohura}
\affiliation{Faculty of Engineering, Yokohama National University, Yokohama 240-8501, Japan }
\author{S.~Ozawa}
\affiliation{National Institute of Information and Communications Technology, Tokyo 184-8795, Japan }
\author{X.~L.~Qian}
\affiliation{Department of Mechanical and Electrical Engineering, Shandong Management University, Jinan 250357, China }
\author{X.~B.~Qu}
\affiliation{College of Science, China University of Petroleum, Qingdao, 266555, China }
\author{T.~Saito}
\affiliation{Tokyo Metropolitan College of Industrial Technology, Tokyo 116-8523, Japan }
\author{M.~Sakata}
\affiliation{Department of Physics, Konan University, Kobe 658-8501, Japan }
\author{T.~K.~Sako}
\affiliation{Institute for Cosmic Ray Research, University of Tokyo, Kashiwa 277-8582, Japan }
\author{J.~Shao}
\affiliation{Key Laboratory of Particle Astrophysics, Institute of High Energy Physics, Chinese Academy of Sciences, Beijing 100049, China }
\affiliation{Institute of Frontier and Interdisciplinary Science and Key Laboratory of Particle Physics and Particle Irradiation (MOE), Shandong University, Qingdao 266237, China }
\author{M.~Shibata}
\affiliation{Faculty of Engineering, Yokohama National University, Yokohama 240-8501, Japan }
\author{A.~Shiomi}
\affiliation{College of Industrial Technology, Nihon University, Narashino 275-8575, Japan }
\author{H.~Sugimoto}
\affiliation{Shonan Institute of Technology, Fujisawa 251-8511, Japan }
\author{W.~Takano}
\affiliation{Faculty of Engineering, Kanagawa University, Yokohama 221-8686, Japan }
\author{M.~Takita$^{\P}$}
\affiliation{Institute for Cosmic Ray Research, University of Tokyo, Kashiwa 277-8582, Japan }
\author{Y.~H.~Tan}
\affiliation{Key Laboratory of Particle Astrophysics, Institute of High Energy Physics, Chinese Academy of Sciences, Beijing 100049, China }
\author{N.~Tateyama}
\affiliation{Faculty of Engineering, Kanagawa University, Yokohama 221-8686, Japan }
\author{S.~Torii}
\affiliation{Research Institute for Science and Engineering, Waseda University, Tokyo 169-8555, Japan }
\author{H.~Tsuchiya}
\affiliation{Japan Atomic Energy Agency, Tokai-mura 319-1195, Japan }
\author{S.~Udo}
\affiliation{Faculty of Engineering, Kanagawa University, Yokohama 221-8686, Japan }
\author{H.~Wang}
\affiliation{Key Laboratory of Particle Astrophysics, Institute of High Energy Physics, Chinese Academy of Sciences, Beijing 100049, China }
\author{H.~R.~Wu}
\affiliation{Key Laboratory of Particle Astrophysics, Institute of High Energy Physics, Chinese Academy of Sciences, Beijing 100049, China }
\author{L.~Xue}
\affiliation{Institute of Frontier and Interdisciplinary Science and Key Laboratory of Particle Physics and Particle Irradiation (MOE), Shandong University, Qingdao 266237, China }
\author{Y.~Yamamoto$^{\dagger}$}
\affiliation{Department of Physics, Konan University, Kobe 658-8501, Japan }
\altaffiliation{\deceased}
\author{Z.~Yang}
\affiliation{Key Laboratory of Particle Astrophysics, Institute of High Energy Physics, Chinese Academy of Sciences, Beijing 100049, China }
\author{Y.~Yokoe}
\affiliation{Institute for Cosmic Ray Research, University of Tokyo, Kashiwa 277-8582, Japan }
\author{A.~F.~Yuan}
\affiliation{Physics Department of Science School, Tibet University, Lhasa 850000, China }
\author{L.~M.~Zhai}
\affiliation{National Astronomical Observatories, Chinese Academy of Sciences, Beijing 100012, China }
\author{H.~M.~Zhang}
\affiliation{Key Laboratory of Particle Astrophysics, Institute of High Energy Physics, Chinese Academy of Sciences, Beijing 100049, China }
\author{J.~L.~Zhang}
\affiliation{Key Laboratory of Particle Astrophysics, Institute of High Energy Physics, Chinese Academy of Sciences, Beijing 100049, China }
\author{X.~Zhang}
\affiliation{School of Astronomy and Space Science, Nanjing University, Nanjing 210093, China }
\author{X.~Y.~Zhang}
\affiliation{Institute of Frontier and Interdisciplinary Science and Key Laboratory of Particle Physics and Particle Irradiation (MOE), Shandong University, Qingdao 266237, China }
\author{Y.~Zhang}
\affiliation{Key Laboratory of Particle Astrophysics, Institute of High Energy Physics, Chinese Academy of Sciences, Beijing 100049, China }
\author{Yi~Zhang}
\affiliation{Key Laboratory of Dark Matter and Space Astronomy, Purple Mountain Observatory, Chinese Academy of Sciences, Nanjing 210034, China }
\author{Ying~Zhang}
\affiliation{Key Laboratory of Particle Astrophysics, Institute of High Energy Physics, Chinese Academy of Sciences, Beijing 100049, China }
\author{S.~P.~Zhao}
\affiliation{Key Laboratory of Particle Astrophysics, Institute of High Energy Physics, Chinese Academy of Sciences, Beijing 100049, China }
\author{Zhaxisangzhu}
\affiliation{Physics Department of Science School, Tibet University, Lhasa 850000, China }
\author{X.~X.~Zhou}
\affiliation{Institute of Modern Physics, SouthWest Jiaotong University, Chengdu 610031, China }
\collaboration{The Tibet AS$\gamma$ Collaboration}

\date{\today}

\begin{abstract}
We report, for the first time, the long-awaited detection of diffuse gamma rays with energies between 100~TeV and 1~PeV in the Galactic disk. Particularly, all gamma rays above 398~TeV are observed apart from known TeV gamma-ray sources and compatible with expectations from the hadronic emission scenario in which gamma rays originate from the decay of $\pi^0$'s produced through the interaction of protons with the interstellar medium in the Galaxy. This is strong evidence that cosmic rays are accelerated beyond PeV energies in our Galaxy and spread over the Galactic disk.
\end{abstract}

\maketitle


\section{Introduction}

Cosmic-ray energy spectrum has approximately a power-law shape $dN/dE
\propto E^p$ in an energy region between 10$^{10}$ eV and 10$^{20}$ eV
\cite{Horandel03}. One of the most prominent features of the spectrum
is the so-called knee at 4$\times$10$^{15}$ eV ($=4$~PeV), where the
spectrum steepens with its power-law index changing from $p=-2.7$ to
$-3.1$ \cite{Kulikov58,Amenomori08}. In a scenario most widely
accepted, cosmic rays are accelerated up to PeV energies by energetic
objects in our Galaxy, such as supernova remnants (SNRs), and well
confined in the Galaxy up to the knee energy by the Galactic magnetic
field \cite{Berezhko99,Kobayakawa02}, although source objects and
acceleration mechanisms are still under discussion. To confirm the
theory predicting the Galactic origin of the PeV cosmic rays,
therefore, it would be conclusive to experimentally identify the
objects in our Galaxy, called ``PeVatrons'', which are accelerating
cosmic rays up to PeV energies.

In recent decades, high-energy gamma-ray observations have been
utilized to identify cosmic-ray sources by detecting the arrival
direction of gamma rays produced by cosmic rays, because gamma rays
travel straight from the source free from the magnetic
deflection. Through the hadronic interaction with ambient matters, PeV
cosmic rays produce neutral pions which decay into gamma rays with
energies as high as 100~TeV \cite{Kelner06,Kappes07,Lipari18}.           
Recently, the ground-based Cherenkov 
telescopes and air shower (AS) arrays observed gamma rays with
energies up to a few tens of TeV from more than 100 sources in the
Galaxy \cite{TeVCat20,Abdalla18,Abeysekara17}. The Tibet AS$\gamma$
\cite{Amenomori19} and HAWC experiments \cite{Abeysekara20} also
detected gamma rays beyond 100~TeV from a few sources which are very
good candidates of PeVatrons. However, there has been no conclusive
evidence for the cosmic-ray PeVatron reported so far.

Gamma-ray telescopes on board satellites, such as the EGRET and ${\it
  Fermi}$-LAT, have precisely observed diffuse gamma rays from the
Galactic disk in an energy range $0.1<E<100$~GeV
\cite{Hunter97,Ackermann12}. The gamma-ray distribution is extended
more than a few degrees in Galactic latitude, similar to the                  
distribution of interstellar gas. The measured spectra are now
established to be dominated by emissions from the interaction of
cosmic rays including electrons with interstellar gas and magnetic
field in this energy region \cite{Ackermann12}. In the higher energy range, the Milagro
experiment reported TeV diffuse gamma-ray emissions from the Cygnus
region in the Galactic disk \cite{Atkins05}, while the ARGO-YBJ
experiment reported diffuse gamma rays with $0.35<E<2$~TeV extended
over Galactic longitude ($l$) between $25^{\circ}<l<100^{\circ}$
\cite{Bartoli15}. Overall, their observed fluxes are consistent with
the standard ${\it Fermi}$-LAT model for the diffuse Galactic
emission. At the highest energy region, the CASA-MIA experiment
presented the upper limits of Galactic diffuse gamma rays with
$140~{\rm TeV}<E<1.3$~PeV \cite{Borione98}.

In this Letter, we report on the detection of diffuse gamma rays with
$100~{\rm TeV}<E<1$~PeV from the Galactic disk with the Tibet air
shower array and muon detector array (Tibet AS+MD array) and present
evidence for PeV cosmic rays being accelerated and confined in the
Galaxy.

\section{Experiment}

In order to observe high-energy gamma rays with high sensitivity, we
started a new hybrid experiment using the surface AS array combined
with the underground water-Cherenkov-type muon detector array at Yangbajing
($90.522^\circ$E, $30.102^\circ$N; 4300~m above sea level) in Tibet, China. The AS array,  
covering a large area of 65700~m$^2$, precisely measures the arrival
direction and energy of each primary cosmic ray, while the underground
muon detector array, with a detection area of 3400~m$^2$ beneath the
AS array, measures number of muons in each AS. Because an AS induced
by a gamma-ray contains much less muons than an AS induced by a
primary cosmic ray in the atmosphere, the muon detector array enables
us to efficiently discriminate cosmic-ray background events from
gamma-ray signals \cite{Sako09}. Based on this technique, we
suppressed more than 99.9\% of cosmic-ray background events above
100~TeV and succeeded in detecting unprecedentedly high-energy gamma
rays from the Crab Nebula. For more details, please see
\cite{Amenomori19}.

\section{Data Analysis}

\begin{figure}[t]
\includegraphics[width=8.5cm]{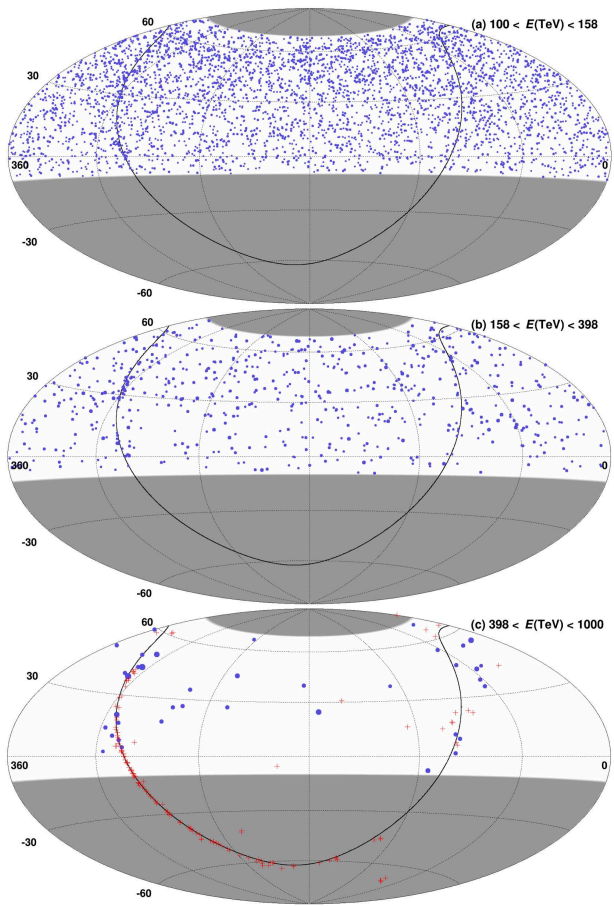}
\caption{ The arrival direction of each gamma-ray-like event observed
  with (a) $100<E<158$~TeV, (b) $158<E<398$~TeV, and (c) $398<E<1000$~TeV, 
  respectively, in the equatorial coordinate.  The blue solid circles
  show arrival directions of gamma-ray-like events observed by the
  Tibet AS+MD array. The area of each circle is proportional to the
  measured energy of each event. The red plus marks show directions of
  the known Galactic TeV sources (including the unidentified sources)
  listed in the TeV gamma-ray catalog \cite{TeVCat20}. The solid curve
  indicates the Galactic plane, while the shaded areas indicate the
  sky regions outside the field of view of the Tibet AS+MD array.
\label{fig_1}}
\end{figure}

The energy and arrival direction of each gamma ray are reconstructed
using the AS particle density and timing recorded at each
scintillation detector composing the AS array. The angular resolution
(50\% containment) is estimated to be approximately $0.22^{\circ}$ and  
0.16$^{\circ}$ for 100~and 400~TeV gamma rays, respectively.
The pointing accuracy has been estimated to be less than $0.06^{\circ}$  
from the observation of the Crab Nebula as described in the Supplemental Material of our previous Letter \cite{Amenomori19}.

To estimate the gamma-ray energy, we use S50 defined as the particle
density detected in an AS surface detector ($\rho$) at a perpendicular
distance of 50~m from the AS axis in the best-fit NKG function
\cite{Kawata17}. The energy resolutions with S50 are roughly estimated
to be 20\% and 10\% for 100~TeV and 400~TeV, respectively. The
absolute energy scale uncertainty was estimated to be 12\% from the
westward displacement of the Moon's shadow center due to the
geomagnetic field \cite{Amenomori09}. The live time of the dataset is
719 days from February 2014 to May 2017, and the average effective detection time for   
the Galactic plane observation is approximately 3700 hours at the zenith angle less than $40^{\circ}$.
The data selection criteria are the same in our previous work \cite{Amenomori19} except for the muon cut condition.
According to the CASA-MIA experiment, the
marginal excess along the Galactic plane in the sub-PeV energies is
1.63$\sigma$ and the fraction of excess to cosmic-ray background
events is estimated to be approximately $3\times10^{-5}$
\cite{Borione98}. In order to search for signals with such a small
excess fraction, we adopt a tight muon cut in the present analyses
requiring for gamma-ray-like events to satisfy $\Sigma N_{\rm \mu} <
2.1\times10^{-4}$ $(\Sigma\rho)^{1.2}$ or $\Sigma N_{\rm \mu}<0.4$,   
where $\Sigma N_{\rm \mu}$ is the total number of muons detected in
the underground muon detector array. This is just one order of
magnitude tighter than the criterion used in our previous work
\cite{Amenomori19}. The cosmic-ray survival ratio with this tight muon
cut is experimentally estimated to be approximately $10^{-6}$ above 400~TeV,
while the gamma-ray survival ratio is estimated to be 30\% by the MC
simulation. The comparison between the cosmic-ray data and the MC simulation
is described in Fig.~S1 in Supplemental Material \cite{SM}.   

\begin{figure}[t]
\includegraphics[width=8.5cm]{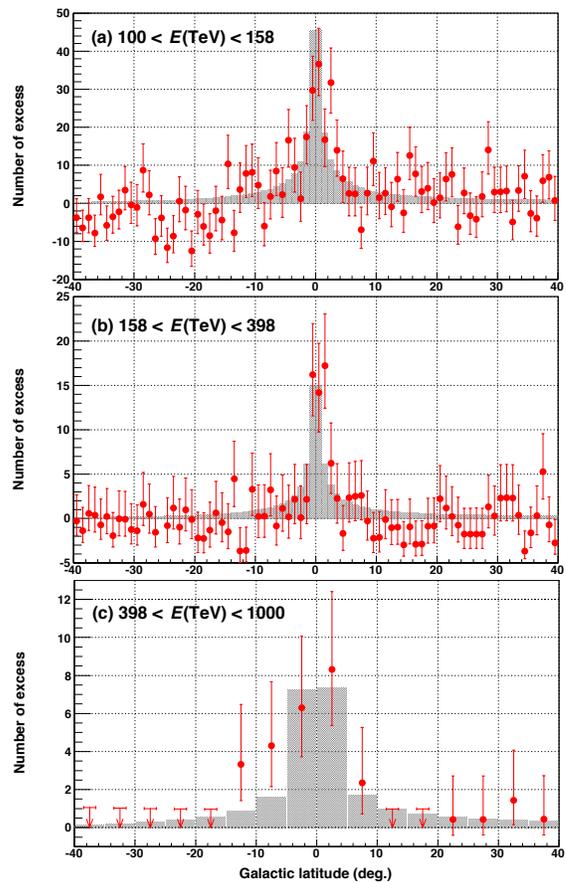}
\caption{ Gamma-ray excess counts as a function of Galactic
  latitude with (a) $100<E<158$~TeV, (b) $158<E<398$~TeV, and (c) 
  $398<E<1000$~TeV.  The excess count is calculated from the observed
  event number after subtracting the estimated background event number
  (see text). The Galactic longitude of the arrival direction in each
  figure is integrated across our FOV (approximately
  $22^{\circ}<l<225^{\circ}$).  The solid circles show the
  experimental data, while the shaded histograms display the model
  profile \cite{Lipari18} rebinned in every (a)(b) $1^{\circ}$ and (c)
  $5^{\circ}$ of the Galactic latitude. The downward arrows show upper  
  limits of excess at 68\% confidence level. The number of excess in
  the model, which is independent of energy, is normalized to the
  observed number within $|b|<5^{\circ}$.
\label{fig_2}}
\end{figure}

\section{Results and Discussion}

Figure~\ref{fig_1} shows arrival directions of gamma-ray-like events
in (a) $100 \ (=10^{2.0}) <E<158 \ (=10^{2.2})$~TeV, (b) $158 \ (=10^{2.2}) <E<398 \ (=10^{2.6})$~TeV and (c) $398         
\ (=10^{2.6}) <E<1000 \ (=10^{3.0})$~TeV, remaining after the tight muon
cut.  It is seen that the observed arrival directions concentrate in a
region along the Galactic plane (see also
Fig.~\ref{fig_2}). Particularly in Fig.~\ref{fig_1} (c), 23        
gamma-ray-like events are observed in $|b|<10^{\circ}$ which we define
as the ON region ($N_{\rm ON}=23$), while only 10 events are observed
in $|b|>20^{\circ}$ which we define as the OFF region ($N_{\rm OFF} =
10$).
Since the total number of events before the tight muon cut is $8.6\times10^{6}$,           
the cosmic-ray survival ratio is estimated to be $1.2\times10^{-6}$ in $|b|>20^{\circ}$ above 398~TeV.
We use $N_{\rm OFF}$ in $|b|>20^{\circ}$ to estimate the number of cosmic-ray
background events, because the contribution from extragalactic gamma
rays in $E>100$~TeV is expected to be strongly suppressed due to the
pair-production interaction with the extragalactic background light.
The mean free path lengths for the pair-production for 100~TeV and 1~PeV are a few Mpc and 10~kpc, respectively \cite{Protheroe00}. 

Since the ratio ($\alpha$) of exposures in ON and OFF regions is
estimated to be $0.27$ by the MC simulation with our geometrical
exposure, the expected number of background events in the ON region
with $|b|<10^{\circ}$ is $N_{\rm BG} = \alpha N_{\rm OFF} = 2.73$ and the
Li-Ma significance \cite{Li83} of the diffuse gamma rays in the ON
region is calculated to be 5.9$\sigma$. The number of events and the                   
significances in each energy bin are summarized in Table~S1 in
Supplemental Material \cite{SM}.
The observed distribution of the number of muons for $E>398$~TeV after the muon cut                     
is consistent with that estimated from the gamma-ray MC simulation as shown in Fig.~S2 in Supplemental Material \cite{SM}.
The highest energy 957($^{+166}_{-141}$)~TeV gamma ray is observed near the Galactic
plane, where the uncertainty in energy is defined as the quadratic sum of the absolute energy-scale error (12\%) and the energy resolution \cite{Amenomori19}.  
Solid circles in Fig.~\ref{fig_2} display $N_{\rm ON}-N_{\rm
  OFF}$ as a function of $b$ in (a) $100<E<158$~TeV, (b) $158<E<398$~TeV, and             
(c) $398<E<1000$~TeV. The concentration of diffuse gamma rays around the
Galactic plane is apparent particularly in Fig.~\ref{fig_2}.

\begin{figure}[t]
\includegraphics[width=8.5cm]{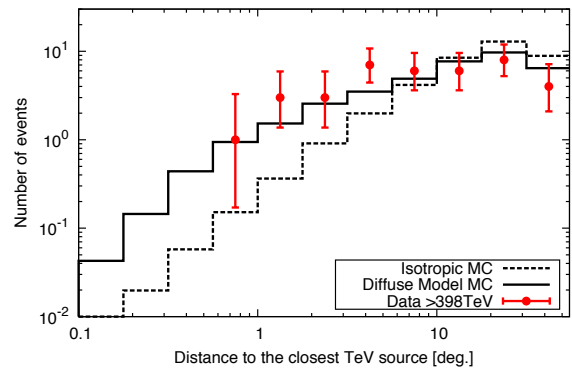}
\caption{ The distribution of the angular distance between the arrival
  direction of each observed gamma-ray-like event with $E>398$~TeV and
  the direction of its closest known TeV source listed in the TeV
  gamma-ray catalog \cite{TeVCat20}. The red solid circles show the
  observed data, while the dashed and solid histograms display the MC
  results expected from the isotropic event distribution and the
  diffuse gamma-ray model \cite{Lipari18}, respectively, to be
  observed with our geometrical exposure.
\label{fig_3}}
\end{figure}

In order to estimate contribution from the known gamma-ray sources, we
searched for gamma-ray signals above 100~TeV from the direction of the
selected 60 Galactic sources (excluding the extragalactic-type
sources, but including the unidentified (UNID) sources) listed in the
TeV source catalog \cite{TeVCat20} within $|b|<5^{\circ}$ in our field
of view (FOV). We used a search window with a radius of $0.5^{\circ}$
centered at each source direction, which contains more than 90\% of
gamma-ray events, as
shown in Fig.~S3 in Supplemental Material \cite{SM}.  Since the source            
extensions of the HAWC sources above 56~TeV were typically around
0.3$^{\circ}$ \cite{Abeysekara20}, the search window radius
0.5$^{\circ}$ is appropriate to exclude most of the contributions from
such extended sources to diffuse gamma rays.  Stacking 60 sources, we
found 37 gamma-ray-like events within search windows against 8.7
background events, which corresponds to 6.8$\sigma$ above 100~TeV,
while the number of all excess within $|b|<5^{\circ}$ ($N_{\rm
  excess}$) is 253.5. The fractional source contribution ($N_{\rm
  point}=37-8.7=28.3$) to the diffuse component ($N_{\rm
  diffuse}=N_{\rm excess}-N_{\rm point}=225.2$) is estimated to be
13\% above 100~TeV.

We also searched for gamma-ray signals within a search window centered
at each direction of 38 gamma-ray-like events in $E>398$~TeV, but we
found no significant signal above 10~TeV. This implies that these 38
events are orphan gamma rays as is expected from the diffuse gamma-ray
scenario, although the existence of unknown sporadic/weak steady
sources with very hard spectra in each direction cannot be ruled out.

Figure~\ref{fig_3} shows the distribution of angular distance between
each of 38 gamma-ray-like events in $E>398$~TeV and its closest
Galactic TeV source.  Surprisingly, there is no gamma-ray excess near
the know TeV sources.  Such high-energy gamma rays which originate
from PeV electrons should be produced near by the sources, due to
significant energy loss via the synchrotron radiation in the magnetic
field around the source.  The observed gamma rays are, therefore, hard
to interpret in the leptonic scenario.  The gamma-ray emission by
electrons will be also significantly suppressed above 100~TeV due to
rapid decrease of Inverse-Compton (IC) cross section by the Klein-Nishina
effect.

Recently, Lipari and Vernetto \cite{Lipari18} developed a model
capable of successfully reproducing the diffuse gamma-ray/neutrino
flux observed in $0.1~{\rm GeV}<E<10$~PeV, by utilizing relevant cosmic-ray
nuclei and electron spectra, interstellar gas distribution, soft
photon field, gamma-ray/neutrino production processes and absorption
effects in the Galaxy. They tested two different models named the
space-independent and space-dependent models. The cosmic-ray spectrum
in the first model is assumed to be identical everywhere in the
Galaxy, while the spectrum in the second model is assumed to be harder
in the central region of the Galaxy than that at the Earth as
indicated by the observed spectral index of Galactic diffuse
gamma rays in $0.1<E<100$~GeV. 
This kind of scenario was also discussed elsewhere \cite{Guo18}.
Both models can reproduce the
observed flux and spatial distribution of arrival directions by ${\it Fermi}$-LAT 
in the GeV energy region. 
The predicted gamma-ray spectrum above
1~GeV is also dominated by the contribution from the hadronic
interaction between the interstellar matter and cosmic rays.
It was concluded that the contribution to the diffuse gamma rays from      
the IC scattering and bremsstrahlung by relativistic electrons is less
than 5\% compared with the hadronic process above 100~TeV, considering
the steep electron and positron spectra with $p=-3.8$ measured by
HESS \cite{Aharonian08}, DAMPE \cite{Ambrosi17} and CALET \cite{Adriani18}.
Another model \cite{Gaggero15} showed the IC scattering contribution
in the low Galactic latitude is negligible above 20~TeV.

Gray histograms in Fig.~\ref{fig_2} show the prediction of the
space-independent model \cite{Lipari18}. It is seen that the
distribution in (a)(b) is overall consistent with the model
prediction. The distribution in (c) observed in $398<E<1000$~TeV looks
broader than that in (a)(b), but it is also statistically consistent
with the prediction rebinned in every $5^{\circ}$ of the Galactic
latitude ($b$).

\begin{figure}[t]
\includegraphics[width=8.5cm]{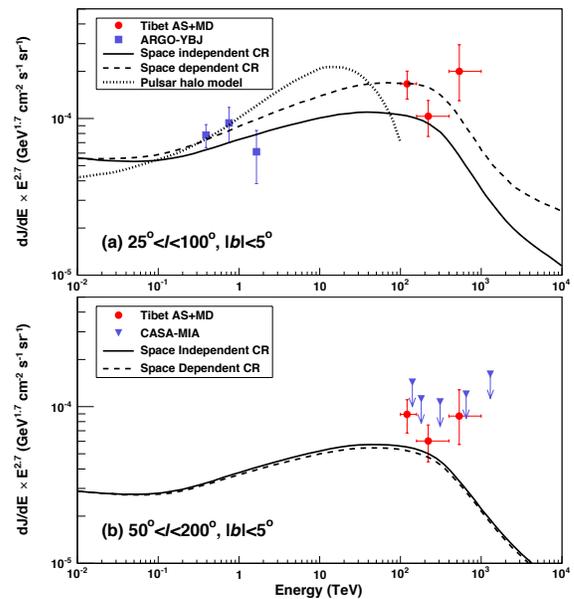}
\caption{ Differential energy spectra of the diffuse gamma rays from
  the Galactic plane in the regions of (a) $|b|<5^{\circ}$,
  $25^{\circ}<l<100^{\circ}$ and (b) $|b|<5^{\circ}$,
  $50^{\circ}<l<200^{\circ}$, respectively.  The solid circles show
  the observed flux after excluding the contribution from the known
  TeV sources listed in the TeV gamma-ray catalog \cite{TeVCat20},
  while the solid and dashed curves display the predicted energy
  spectra by the space-independent and space-dependent models by
  Lipari and Vernetto \cite{Lipari18}, respectively (see text).
  The dotted curve in panel (a) shows the flux predicted by a leptonic model \cite{Linden18}
  in which gamma rays are induced by relativistic electrons and positrons from pulsars. 
  Solid squares in panel (a) and triangles with arrows in panel (b) indicate
  the flux measured by ARGO-YBJ \cite{Bartoli15} and the flux upper limit by the
  CASA-MIA experiment \cite{Borione98}, respectively. The error bar shows 1$\sigma$ 
  statistical error.
\label{fig_4}}
\end{figure}

Figure~\ref{fig_4} shows the observed differential energy spectra of
diffuse gamma rays, compared with the model predictions by Lipari and
Vernetto \cite{Lipari18} in which gamma-ray spectra are calculated in
(a) $25^{\circ}<l<100^{\circ}$ and (b) $50^{\circ}<l<200^{\circ}$
along the Galactic plane, each in $|b|<5^{\circ}$.
The measured fluxes by the Tibet AS+MD array are summarized in Table~S2 in Supplemental Material \cite{SM}.   
These fluxes are obtained after subtracting
events within $0.5^{\circ}$ from the known TeV sources and the
systematic error of the observed flux is approximately 30\% due to the
uncertainty of absolute energy-scale \cite{Amenomori09}.  We corrected
time variation of detector gain at each detector based on the single
particle measurement for each run. The time variation of gamma-ray-like
excess above 100~TeV in $|b|<5^{\circ}$ is stable within
approximately 10\%.  It is seen that the measured fluxes by the Tibet
AS+MD array are compatible with both the space-independent and
space-dependent models based on the hadronic scenario.
As a leptonic model, it is proposed that gamma-ray halos induced by the            
relativistic electrons and positrons from pulsars explain the
Galactic diffuse gamma rays above 500~GeV \cite{Linden18}.
However, the gamma-ray flux predicted by this model has an exponential cutoff well below 100~TeV, and
is inconsistent with the observation by Tibet AS+MD array (see Fig.~\ref{fig_4}(a)).

The observed flux in the highest energy bin in $398<E<1000$~TeV looks higher than
the model prediction, but it is not inconsistent with the model when
the statistical and systematic errors are considered. Above 398~TeV,
the total number of observed events is 10 in each of
$25^{\circ}<l<100^{\circ}$ and $50^{\circ}<l<200^{\circ}$, which
includes the Cygnus region around $l=80^{\circ}$. Interestingly, 4 out
of 10 events are detected within $4^{\circ}$ from the center of the
Cygnus cocoon, which is claimed as an extended gamma-ray source by the
ARGO-YBJ \cite{Bartoli14} and also proposed as a strong candidate of
the PeVatrons \cite{Aharonian19}, but not taken into account in the
model \cite{Lipari18}. If these 4 events are simply excluded, the
observed flux at the highest energy in Fig.~\ref{fig_4} better
agrees with model predictions.

The high-energy astrophysical neutrinos are also a good probe of the
spectrum and spatial distribution of PeV cosmic rays in the Galaxy
\cite{Albert18,Aartsen19}. According to Lipari and Vernetto
\cite{Lipari18}, the diffuse gamma-ray/neutrino fluxes predicted near
the Galactic center ($|l|<30^{\circ}$) by the space-dependent model are
more than 5 times higher than that predicted by the space-independent
model in $100$~TeV $<E<10$~PeV.
Therefore, the gamma-ray/neutrino observations in the southern hemisphere
will also play important roles to understand or constrain the spatial
distribution of PeV cosmic rays in the Galaxy.
Probing PeV diffuse gamma rays/neutrinos from the large-scale structures,
such the Fermi-bubble \cite{Yang19} and the possible dark matter
halo in the Galaxy \cite{Esmaili13,Murase15}, will be also interesting.

\section{Conclusions}

We successfully observed the Galactic diffuse gamma rays in
$100~{\rm TeV}<E<1$~PeV by the Tibet AS+MD array. Particularly, in the
energy region above 398~TeV, we found 23 gamma-ray-like events against
2.73 background events, which corresponds to 5.9$\sigma$ statistical
significance, in $|b|<10^{\circ}$ in our FOV. 
The highest energy of the observed gamma ray is 957($^{+166}_{-141}$)~TeV, which is nearly 1~PeV. 
The gamma-ray distribution are extended around the Galactic plane apart
from known Galactic TeV gamma-ray sources. 
We also found no significant signal above 10~TeV in directions of 38 gamma-ray-like
events above 398~TeV, which implies that these events are
orphan gamma rays as is expected from the diffuse gamma-ray scenario.
The measured fluxes are overall consistent with recent models assuming
the hadronic cosmic-ray origin. These facts are hard to interpret with
the leptonic cosmic-ray origin, indicating that sub-PeV diffuse gamma
rays are produced by the hadronic interaction of protons, which are
accelerated up to a few PeV energies (or possibly $\sim10$~PeV) and escaping from the source,
with the interstellar gas in our Galaxy. Hence, we conclude that the
PeVatrons inevitably exist in the present and/or past Galaxy
accelerating cosmic rays which spread in the Galaxy being well
confined around the Galactic disk.

\begin{acknowledgments}
The collaborative experiment of the Tibet Air Shower Arrays has been
conducted under the auspices of the Ministry of Science and Technology
of China and the Ministry of Foreign Affairs of Japan.
This work was supported in part by a Grant-in-Aid for Scientific Research on Priority Areas from the Ministry of Education, Culture, Sports, Science and Technology,
and by Grants-in-Aid for Science Research from the Japan Society for the Promotion of Science in Japan.
This work is supported by the National Key R\&D Program of China (No. 2016YFE0125500),
the Grants from the National Natural Science Foundation of China (Nos. 11533007, 11673041, and 11873065), 
and the Key Laboratory of Particle Astrophysics, Institute of High Energy Physics, CAS.
This work is also supported by the joint research program of the Institute for Cosmic Ray Research (ICRR), the University of Tokyo.
\end{acknowledgments}

\noindent
\\
$^{\dagger}$ Deceased\\
\\
Corresponding authors:\\
$^{*}$ kawata@icrr.u-tokyo.ac.jp\\
$^{\ddagger}$ huangjing@ihep.ac.cn\\
$^{\P}$ takita@icrr.u-tokyo.ac.jp\\
$^{\S}$ chending@bao.ac.cn\\






%

\appendix


\setcounter{table}{2}
\renewcommand{\thetable}{S\arabic{table}}

\begin{table*}[h]
  \caption{\label{tab_3} (Table in the Supplemental Material) Event IDs and arrival directions in the equatorial coordinates (Right Ascension, Declination) of the gamma-ray like events with $398 < E < 1000$~TeV observed by the Tibet AS+MD array during period between February 2014 and May 2017.}
  \begin{tabular}{cdd}
\hline\hline
\multicolumn{1}{c}{\textrm{ \ \ TASG \ \ }} & \multicolumn{1}{c}{\textrm{ \ \ R.A. J2000 \ \ }}  & \multicolumn{1}{c}{\textrm{ \ \ Dec. J2000 \ \ }} \\
\multicolumn{1}{c}{\textrm{ \ \ Event ID \ \ }} & \multicolumn{1}{c}{\textrm{ \ \ (degrees) \ \ }}  & \multicolumn{1}{c}{\textrm{ \ \ (degrees) \ \ }} \\  \hline
TASG-D01-001  &      18.74  &    55.31 \\
TASG-D01-002  &      26.44  &    68.23 \\
TASG-D01-003  &      35.21  &    54.46 \\
TASG-D01-004  &      49.16  &    44.38 \\
TASG-D01-005  &      55.90  &    43.25 \\
TASG-D01-006  &      62.31  &    38.11 \\
TASG-D01-007  &      63.13  &    55.26 \\
TASG-D01-008  &      63.72  &    34.74 \\
TASG-D01-009  &      67.01  &    46.54 \\
TASG-D01-010  &      96.16  &     9.02 \\
TASG-D01-011  &      98.31  &    11.21 \\
TASG-D01-012  &      99.60  &     1.58 \\
TASG-D01-013  &     114.74  &    -7.55 \\
TASG-D01-014  &     127.01  &    38.26 \\
TASG-D01-015  &     174.45  &    24.48 \\
TASG-D01-016  &     183.43  &    39.60 \\
TASG-D01-017  &     228.12  &    26.53 \\
TASG-D01-018  &     230.56  &    44.40 \\
TASG-D01-019  &     243.22  &    66.27 \\
TASG-D01-020  &     255.47  &    26.46 \\
TASG-D01-021  &     256.49  &    35.31 \\
TASG-D01-022  &     261.10  &    25.56 \\
TASG-D01-023  &     264.29  &    17.95 \\
TASG-D01-024  &     284.38  &     4.50 \\
TASG-D01-025  &     286.96  &     7.96 \\
TASG-D01-026  &     290.28  &    16.36 \\
TASG-D01-027  &     291.45  &    10.03 \\
TASG-D01-028  &     293.62  &    20.36 \\
TASG-D01-029  &     295.63  &     2.30 \\
TASG-D01-030  &     297.17  &    13.82 \\
TASG-D01-031  &     305.44  &    44.21 \\
TASG-D01-032  &     307.08  &    39.02 \\
TASG-D01-033  &     308.69  &    43.92 \\
TASG-D01-034  &     309.49  &    51.05 \\
TASG-D01-035  &     312.33  &    40.23 \\
TASG-D01-036  &     320.32  &    49.46 \\
TASG-D01-037  &     354.97  &    49.65 \\
TASG-D01-038  &     359.96  &    59.19 \\
\hline\hline
\end{tabular}
\end{table*}

\end{document}